\documentclass[final,journal,compsoc]{IEEEtran}

 \usepackage{tikz}
 \usetikzlibrary{arrows,chains,matrix,positioning,scopes,shapes.geometric,decorations.pathmorphing,calc,patterns}
\usepackage[utf8]{inputenc}
\usepackage[english, british]{babel}

\usepackage[binary-units]{siunitx}

\usepackage{hyperref}

\hypersetup{
  pdfauthor={Sebastian E. Schmittner},
  pdftitle={A SAT-based Public Key Cryptography Scheme}
}

\usepackage{yfonts}

\usepackage{enumitem}

\usepackage[cmex10]{amsmath}

\usepackage{amssymb,bbm}
\allowdisplaybreaks[1]

 \DeclareFontFamily{OT1}{pzc}{}
 \DeclareFontShape{OT1}{pzc}{m}{it}{<-> s * [1.10] pzcmi7t}{}
 \DeclareMathAlphabet{\mathpzc}{OT1}{pzc}{m}{it}

 \newcommand{\priv}{\mathpzc{priv}}
 \newcommand{\pub}{\mathpzc{pub}}

\title{A SAT-based Public Key Cryptography Scheme}

\author{Sebastian  E.~Schmittner%}
 \IEEEcompsocitemizethanks{\IEEEcompsocthanksitem
  Universit\"at zu K\"oln\par
  Institut f\"ur Theoretische Physik\par
  Z\"ulpicher Stra\ss{}e~77;
  D-50937~K\"oln\par
 \href{mailto:ses@thp.uni-koeln.de}{ses@thp.uni-koeln.de}
  }}

 \IEEEtitleabstractindextext{%
 \begin{abstract}
   A homomorphic public key crypto-scheme based on the Boolean
   Satisfiability Problem is proposed. The public key is a SAT formula satisfied by
   the private key. Probabilistic encryption generates
   functions implied to be false by the public key XOR the message
   bits. A zero-knowledge proof is used to provide signatures.
 \end{abstract}
\begin{IEEEkeywords}
Public key cryptosystems, Cryptographic protocols, Data Encryption,
Message authentication, Digital signatures.
\end{IEEEkeywords}
 }

\begin{document}

\maketitle

\section{Introduction\label{sec:introduction}}
%ROFL! really? 
%\IEEEPARstart{A}{symmetric}
%why not use this one: :D
\yinipar{U}\textsc{nlike} the symmertic ones,
asymmetric cryptography schemes
predominantly used today are vulnerable (at least) to attacks from
quantum computers using Shor's algorithm. Assuming P$\neq$NP and that NP-hard problems can not be
solved efficiently, not even on a quantum computer, cryptography based
on NP-hard problems is dubbed ``post-quantum''.
 Daniel Bernstein
\cite{BBD09} lists Hash-based, Code-based,
Lattice-based, and Multivariate-quadratic-equations
cryptography as the existing post-quantum algorithms.  
The aim of this
paper is to introduce a different crypto-system based on
the Boolean Satisfiability Problem (SAT). This problem of finding a
pre-image of $1$ under a Boolean function given in a certain
conjunctive form (see Section~\ref{sec:single-key}) is well known to be NP-complete. Although a SAT
instance can, from a certain angle, be viewed as a
multivariate-equation, our crypto-scheme is different from the
systems mentioned above. In particular, we use a fulfilling assignment
as the secret key rather than a trap door. 

Using SAT to provide a post-quantum key pair is not a
far-fetched idea. The main point of this paper is how to use them to
encrypt and decrypt. To this end, we randomly produce
Boolean functions which are implied to be false by the public key and hence
evaluate to $0$ on the private key. These functions $\oplus$ the
message bits form the ciphertext.

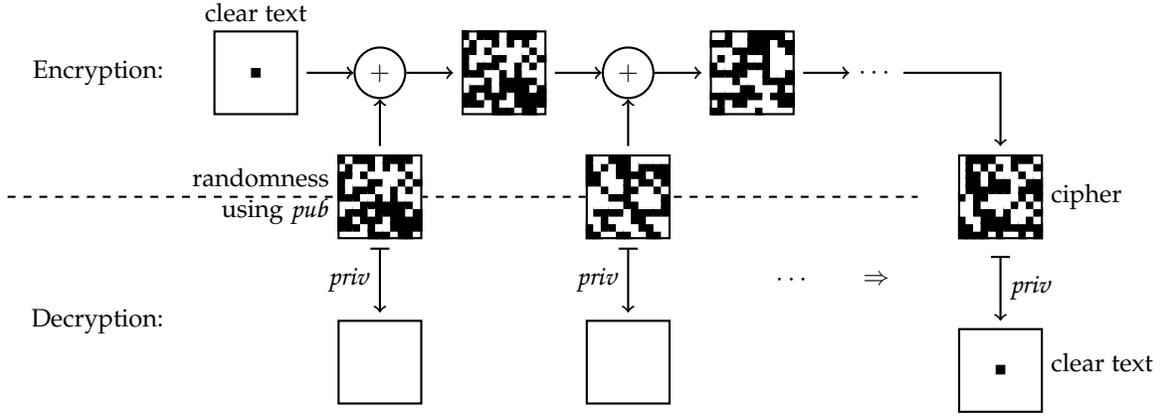
\begin{figure*}
\centering
% \documentclass[a4paper]{amsart}
% \usepackage[utf8]{inputenc}
% \usepackage[english, british]{babel}

% \usepackage[binary-units]{siunitx}

% \usepackage{tikz}
% \usetikzlibrary{arrows,chains,matrix,positioning,scopes,shapes.geometric,decorations.pathmorphing,calc,patterns}
% \usetikzlibrary{decorations.markings}

% \usetikzlibrary{external}
% \tikzexternalize

% \begin{document}

\definecolor{color1}{rgb}{1,1,1}
%\definecolor{color-1}{rgb}{0.05,0.3,0}
\definecolor{color-1}{rgb}{0,0,0}

\pgfmathsetlengthmacro{\pxl}{0.1cm}
\pgfmathsetlengthmacro{\total}{1.1cm}
\pgfmathtruncatemacro{\L}{\total/\pxl+1}
\pgfmathtruncatemacro{\center}{\L/2}
\pgfmathtruncatemacro{\centerm}{\center +1}

\begin{tikzpicture}[thick]

\filldraw[color-1](\centerm*\pxl,\centerm*\pxl) rectangle (\center*\pxl, \center*\pxl);

\node[above] at (0.5*\total,\total) {clear text};
\draw (0,0) rectangle (\total,\total);

\draw[dashed](-2.5*\total,-\total) -- (8.5*\total,-\total);

\node[left] at (-0.5*\total,0.5*\total){Encryption:};
\node[left] at (-0.5*\total,-2.5*\total){Decryption:};

\node(clearText) at (\total,0.5*\total){};

\begin{scope}[yshift=-1.5*\total, xshift=1.5*\total]
\pgfmathsetseed{7}
 \foreach \x in {1,...,\L}{
   \foreach \y in {1,...,\L}{
     \pgfmathtruncatemacro{\xm}{\x-1}
     \pgfmathtruncatemacro{\ym}{\y-1}
     \pgfmathtruncatemacro{\R}{rnd*2}
     \pgfmathtruncatemacro{\r}{rnd*2}

     \pgfmathtruncatemacro{\C}{2*\R-1}
     \filldraw[color\C](\xm*\pxl,\ym*\pxl) rectangle (\x*\pxl, \y*\pxl);
   }
 }
\draw (0,0) rectangle (\total,\total);
\node (rand1) at (0.5*\total,\total) {};
\node (rand1b) at (0.5*\total,0) {};
\node[left] at (0,0.5*\total){\begin{minipage}{4cm}\raggedleft
    randomness\\
using $\pub$\end{minipage}};
\end{scope}

\begin{scope}[ xshift=3*\total]
\pgfmathsetseed{7}
 \foreach \x in {1,...,\L}{
   \foreach \y in {1,...,\L}{
     \pgfmathtruncatemacro{\xm}{\x-1}
     \pgfmathtruncatemacro{\ym}{\y-1}
     \pgfmathtruncatemacro{\R}{rnd*2}
     \pgfmathtruncatemacro{\r}{rnd*2}

     \pgfmathtruncatemacro{\C}{(2*\R-1) * (-2)*((\x == \center)*(\y ==
       \center) -0.5)}

     \filldraw[color\C](\xm*\pxl,\ym*\pxl) rectangle (\x*\pxl, \y*\pxl);
   }
 }
\draw (0,0) rectangle (\total,\total);

\node (summed1) at (0,0.5*\total) {};
\node (summed1r) at (\total,0.5*\total) {};
\end{scope}

\node[circle, draw] (sum1) at (2*\total,0.5*\total){$+$};

\draw (clearText) edge[->] (sum1)
(rand1) edge[->] (sum1)
(sum1) edge[->] (summed1);

\begin{scope}[yshift=-1.5*\total, xshift=4.5*\total]
\pgfmathsetseed{7}
 \foreach \x in {1,...,\L}{
   \foreach \y in {1,...,\L}{
     \pgfmathtruncatemacro{\xm}{\x-1}
     \pgfmathtruncatemacro{\ym}{\y-1}
     \pgfmathtruncatemacro{\R}{rnd*2}
     \pgfmathtruncatemacro{\r}{rnd*2}

     \pgfmathtruncatemacro{\C}{2*\r-1}

     \filldraw[color\C](\xm*\pxl,\ym*\pxl) rectangle (\x*\pxl, \y*\pxl);
   }
 }
\draw (0,0) rectangle (\total,\total);

\node (rand2) at (0.5*\total,\total) {};
\node (rand2b) at (0.5*\total,0) {};
\end{scope}

\begin{scope}[ xshift=6*\total]
\pgfmathsetseed{7}
 \foreach \x in {1,...,\L}{
   \foreach \y in {1,...,\L}{
     \pgfmathtruncatemacro{\xm}{\x-1}
     \pgfmathtruncatemacro{\ym}{\y-1}
     \pgfmathtruncatemacro{\R}{rnd*2}
     \pgfmathtruncatemacro{\r}{rnd*2}

     \pgfmathtruncatemacro{\C}{(2*\R-1) * (-2)*((\x == \center)
*(\y==\center) -0.5)* (2*\r-1)}

     \filldraw[color\C](\xm*\pxl,\ym*\pxl) rectangle (\x*\pxl, \y*\pxl);
   }
 }
\draw (0,0) rectangle (\total,\total);

\node (summed2) at (0,0.5*\total) {};
\node (summed2r) at (\total,0.5*\total) {};
\end{scope}

\node[circle, draw] (sum2) at (5*\total,0.5*\total){$+$};
\node (sum3) at (8*\total,0.5*\total){$\cdots$};

\draw (summed1r) edge[->] (sum2)
(rand2) edge[->] (sum2)
(sum2) edge[->] (summed2)
(summed2r) edge[->] (sum3);

\begin{scope}[ xshift=9*\total, yshift=-1.5*\total]
\fill[white] (0,0) rectangle (\total,\total);
\pgfmathsetseed{42}
 \foreach \x in {1,...,\L}{
   \foreach \y in {1,...,\L}{
     \pgfmathtruncatemacro{\xm}{\x-1}
     \pgfmathtruncatemacro{\ym}{\y-1}
     \pgfmathtruncatemacro{\R}{rnd*2}

     \pgfmathtruncatemacro{\C}{(2*\R-1)}

     \filldraw[color\C](\xm*\pxl,\ym*\pxl) rectangle (\x*\pxl, \y*\pxl);
   }
 }
\draw (0,0) rectangle (\total,\total);
\node (cipherT) at (0.5*\total,\total) {};
\node (cipherB) at (0.5*\total,-0.1*\total) {};
\node[right] at (\total,0.5*\total) {cipher};
\end{scope}

\draw[->] (sum3) -- +(1.5*\total,0)-- (cipherT);

%%%%%%%%%%%%%%%%%%%%%%%%

\draw (1.5*\total,-2.5*\total) rectangle +(\total,-\total);
\draw (rand1b) edge[|->] node[midway, left]{$\priv$} +(0,-0.9*\total);

\draw (4.5*\total,-2.5*\total) rectangle +(\total,-\total);
\draw (rand2b) edge[|->] node[midway, left]{$\priv$} +(0,-0.9*\total);

\node at (7*\total,-2*\total){$\cdots$};
\node at (8*\total,-2*\total){$\Rightarrow$};

\node[xshift=-0.5*\total, yshift=-\total] (decrypted) at (cipherB){};
\draw (decrypted) rectangle +(\total,-\total);

\draw[|->] (cipherB) edge[|->] node[midway, right]{$\priv$} +(0,-0.9*\total);

\filldraw[color-1](9*\total + \center*\pxl,-3.5*\total + \center*\pxl)
rectangle +(\pxl,-\pxl);

\node[right] at (10*\total,-3*\total){clear text};

\end{tikzpicture}
\caption{Encryption (top): The text is hidden in random noise patterns
  generated from the public key $\pub$. Subsequent additions hide the
  structure of the noise patterns to make it harder to subtract the
  noise.\newline
Decryption (bottom): Via the private key, every noise pattern generated
   from $\pub$ is ``switched off'' and hence the clear text is
   revealed from the sum.
}
\label{fig:encoding}
\end{figure*}

The paper is organised as follows: We illustrate the main point of
using randomness for encryption in a (somewhat over simplified)
picture in Figure~\ref{fig:encoding}.  The guiding example in
Section~\ref{sec:example} illustrates the main ideas more accurately.
In Section~\ref{sec:single-key}, we describe and discuss our
algorithms in detail. The probabilistic scheme introduced there is
vulnerable to an oracle attack, which is discussed in
Section~\ref{sec:known-attacks-impr}, as well as some other
attacks. In this section we also compute bounds for the parameters of
the encryption to resist all mentioned attacks and also explain how
the oracle attack can be countered.  In Section~\ref{sec:ident-sign}
we introduce an identification and signature scheme, which is
independent from the encryption.  We conclude in
Section~\ref{sec:conclusion}.

\subsection{Example\label{sec:example}}

Before explaining the key generation and encryption algorithm in
detail, we start with a simple toy example to guide the readers
intuition.  A possible\footnote{Of course, the ratio of clauses to variables in
realistic keys needs to be much larger (see Appendix~\ref{sec:benchmarks}) and also the key length needs
to be much longer.} private/public key pair for Alice is:
%single column:
% \begin{align}
%   \priv & = (1,1,0,0,1,0,0)\\
%   \pub & =
% (\bar x_1 \vee \bar x_2 \vee \bar  x_3) \wedge
% ( \bar x_1 \vee  x_4  \vee x_5) \wedge
% ( x_1 \vee  x_6 \vee  x_7)
% \,
% \end{align}
%
%double column 12pt:
\begin{align}
  \priv & = (1,1,0,0,1,0,0)\\
  \pub & =
(\bar x_1 \vee \bar x_2 \vee \bar  x_3) \wedge
( \bar x_1 \vee  x_4  \vee x_5) \\&\quad \wedge
( x_1 \vee  x_6 \vee  x_7)
\,
\end{align}
since $\pub(\priv)=1$. Bob wants to send a bit $y\in\mathbb B$ to
Alice. He rewrites
% \begin{align}
% \bar c_1 &= x_1  x_2  x_3
% \\
% \bar c_2 &= x_1  \oplus x_1 x_4 \oplus x_1 x_5  \oplus x_1x_4x_5 \\
% \bar c_3 &= 1\oplus x_1 \oplus x_6 \oplus x_7 \oplus x_1x_6 \oplus
% x_1x_7 \oplus x_6  x_7 
% \oplus x_1x_6x_7 
% \end{align}
\begin{align}
\bar c_1 &= x_1  x_2  x_3
\\
\bar c_2 &= x_1  \oplus x_1 x_4 \oplus x_1 x_5  \oplus x_1x_4x_5 \\
\bar c_3 &= 1\oplus x_1 \oplus x_6 \oplus x_7 \oplus x_1x_6 \oplus
x_1x_7 \oplus x_6  x_7  \\&\quad \oplus x_1x_6x_7 
\end{align}
and randomly generates
%single
% \begin{align}
% R_{2,3}&= 
% x_4 \oplus x_5 \oplus x_6 \oplus x_4x_5 \oplus x_4x_6x_7 \oplus
% x_5x_6x_7
% \oplus x_4x_5x_6x_7
% \\
% R_{1,3} &=
% 1\oplus x_2x_3 \oplus x_6x_7 \oplus x_2x_3x_6x_7
% \\
% R_{1,2}&= 
% 1\oplus x_4 \oplus x_5 \oplus x_2x_3 \oplus x_4x_5
% \;.
% \end{align}
%
%double
\begin{align}
R_{2,3}&= 
x_4 \oplus x_5 \oplus x_6 \oplus x_4x_5 \oplus x_4x_6x_7 \\&\quad \oplus
x_5x_6x_7
\oplus x_4x_5x_6x_7
\\
R_{1,3} &=
1\oplus x_2x_3 \oplus x_6x_7 \oplus x_2x_3x_6x_7
\\
R_{1,2}&= 
1\oplus x_4 \oplus x_5 \oplus x_2x_3 \oplus x_4x_5
\;.
\end{align}
The ciphertext is
%two column:
\begin{subequations}
\begin{align}
g &= \bar c_1 R_{2,3} \oplus \bar c_2 R_{1,3} \oplus \bar c_3 R_{1,2}
\\&=y \oplus 1 \oplus
x_4 \oplus x_5 \oplus x_6 \oplus x_7 \oplus x_1x_6 \oplus x_1x_7
\\&\quad
\oplus x_2x_3 
\oplus x_4x_5 \oplus x_4x_6 
\oplus x_4x_7 \oplus x_5x_6 \oplus x_5x_7 
\\&\quad
\oplus x_6x_7 \oplus
x_1x_4x_6 \oplus x_1x_4x_7 \oplus x_1x_5x_6 \oplus x_1x_5x_7 
\\&\quad
\oplus x_2x_3x_6 \oplus x_2x_3x_7 \oplus x_4x_5x_6 \oplus x_4x_5x_7
\\&\quad
\oplus x_4x_6x_7 \oplus x_5x_6x_7 \oplus x_1x_2x_3x_7 
\oplus x_1x_4x_5x_6 
\\&\quad
\oplus x_1x_4x_5x_7 \oplus x_2x_3x_6x_7 \oplus x_4x_5x_6x_7
\end{align}
\end{subequations}
%
%one column:
% \begin{subequations}
% \begin{align}
% g &= \bar c_1 R_{2,3} \oplus \bar c_2 R_{1,3} \oplus \bar c_3 R_{1,2}
% \\&=y \oplus 1 \oplus
% x_4 \oplus x_5 \oplus x_6 \oplus x_7 \oplus x_1x_6 \oplus x_1x_7
% \oplus x_2x_3 
% \oplus x_4x_5 \oplus x_4x_6 
% \oplus x_4x_7 \oplus x_5x_6 \oplus x_5x_7 
% \oplus x_6x_7 \oplus
% x_1x_4x_6 
% \\&\quad
% \oplus x_1x_4x_7 \oplus x_1x_5x_6 \oplus x_1x_5x_7 
% \oplus x_2x_3x_6 \oplus x_2x_3x_7 \oplus x_4x_5x_6 \oplus x_4x_5x_7
% \oplus x_4x_6x_7 \oplus x_5x_6x_7 \oplus x_1x_2x_3x_7 
% \\&\quad
% \oplus x_1x_4x_5x_6 
% \oplus x_1x_4x_5x_7 \oplus x_2x_3x_6x_7 \oplus x_4x_5x_6x_7
% \end{align}
% \end{subequations}
Bob sends the ciphertext $g$ for the cleartext $y$ to Alice, who decodes $g(\priv)=y$.

Mallet can attack the scheme above by replacing a literal in
$g$ by a truth value, say $x_3\mapsto 1$. He sends the modified ciphertext to Alice and
from Alice's reply, he will notice whether Alice received the correct
bit or not. Hence he concludes that $x_3=0$ in the private key.  Alice
might notice the attack by a suitable syntax check on the decoded
text (containing more than one bit), but the security of the key pair
is lost anyway. 

This attack can be countered as follows. Bob seeds his random number
generator, which chooses the clauses and generates the random
functions  $R_i$, with a hash of the 
salted clear text. He then sends the salt
together with the encrypted message. If the clear text was long
enough, it would take Mallet very long to guess the right seed from the
salt alone (hence decoding the message). But knowing the salt and the
clear text, Alice can easily re-encrypt the message after decryption
and check whether the ciphertext has been altered by Mallet. She  rejects
any faulty message, regardless of the private key, hence not
revealing any information.

\section{Algorithms\label{sec:single-key}}

\yinipar{G}\textsc{eneration} of key pairs and the 
encryption algorithm are discussed in this section.   The public key, $\pub$, is a
``planted'' random $k$-SAT instance for $k\in \mathbb N$ with
$k>2$. This is a Boolean function in $n \in \mathbb N$ variables,
$\pub:\mathbb B^n \to \mathbb B$, which is given as a conjunction of
$m \in \mathbb N$ many
$k$-clauses, $c_j: \mathbb B^k \subset \mathbb B^n \to \mathbb B$, together with a truth
assignment, the private key $\priv \in \pub^{-1}(1)$.   In summary
\begin{align}
\pub &: x \mapsto \bigwedge_{j=1}^m c_j(x)\;,&c_j &: x \mapsto \bigvee_{i=1}^k (x_{I(i,j)} \oplus s(i,j))\;,
\end{align}
where the signs $s(i,j)\in \mathbb B$ determine wether the literal
$x_I$ is negated.
To discuss the complexity of the algorithms, we use the notation
\begin{align}
O(f)&:= \left\{g:\mathbb N \to \mathbb R \;\big|\;
\limsup_{n\to\infty}\frac{|g(n)|}{f(n)}<\infty \right\} \;,\\
% \Omega(f)&:=\{g:\mathbb N \to \mathbb R^+ \;\big|\;
% \limsup_{n\to\infty}\frac{f(n)}{g(n)}<\infty\}\\
o(f)&:=\left\{g:\mathbb N \to \mathbb R \;\big|\;
\limsup_{n\to\infty}\frac{|g(n)|}{f(n)}=0\right\}\;,
\end{align}
and $\Theta(f):=O(f)\setminus o(f)$ for $f:\mathbb N \to \mathbb R^+$.
% For rougher estimates we will also write
% \begin{equation}
% f \xrightarrow{n\to\infty} g \Leftrightarrow \limsup_{n\to\infty}\frac{|f(n) -
%   g(n)|}{|f(n)|+|g(n)|} = 0\;.
% \end{equation}

\subsection{Key Generation\label{sub:single-keygen}}

A key pair is generated as follows:
\begin{enumerate}
\item Choose the private key $\priv \in \mathbb B^n$ at random.

\item Generate the  clause $c_j$ by randomly choosing $k$
  distinct integers $I_{j,i}\in\{1,\ldots,n\}$ and ``signs'' $s_i \in \mathbb
  B$. 

\item If $c_j(\priv)$ accept the clause, otherwise reject
  it.\footnote{If the algorithm is to have finite worst case run time,
    one should modify the clause instead of generating a new one at
    random. Notice, however, that the number of signs to flip has to
    be chosen carefully if the resulting distribution of public keys
    is to be unchanged. See e.g.~\cite{Krzakala2012}.}

\item Repeat until $m$ clauses have been generated.
\end{enumerate}

In other words, the data to be generated and stored for the each key
pair is
\begin{align}
I&:
\{1,\ldots,m\}\times\{1,\ldots,k\} \to  \{1, \ldots, n\}
\\
\text{and}\quad s &:
\{1,\ldots,m\}\times\{1,\ldots,k\} \to  \mathbb B\;,
\end{align}
hence the length of the public key is in $\Theta(m k (\log(n)+1))$.
The run time is proportional to the public key length\footnote{If
  clauses are rejected this is tha case with probability exponentially
  close to $1$.}.

It is very important to generate hard instances as public
keys. Therefore we must ensure that $\priv$ is (almost surely) the
unique solution, i.e.\ $m$ needs to be larger than the critical value,
$m>m_c=\alpha_c(k) n$ for some $\alpha_c(k)\in\Theta(1)$, see
\cite{Kirkpatrick1994}.  Choosing $m$ rather close to $m_c$ yields the
most difficult instance. This is well known in similar setups, see
Appendix~\ref{sec:benchmarks} for some of our own benchmarks.
Choosing $m \propto n$ for fixed $k$ leads to a public key length of
$\Theta(n\log(n))$. A different scaling of $m$ with $n$ is not
recommended, since the resulting SAT instances then become much easier
to solve.

\subsection{Encryption and Decryption\label{sec:encAlg}}
We fix parameters $\alpha,\beta \in \mathbb N$ with $2\leq \beta \ll
\alpha$ to tune the run time and security of the encryption. 

\begin{enumerate}
\item \label{encryptChoose} Choose $\alpha$ many tuples of $\beta$
  many distinct clauses  at
  random,
%\begin{equation}
$
J : \{1,\ldots,\alpha\}\times\{1,\ldots,\beta\}\to\{1,\ldots,m\} 
$
%\end{equation}
such that $J(i,a)\neq J(i,b)$ for all $a\neq b$.

\item \label{item:17} The ciphertext encoding $ y \in \mathbb B$
  is the algebraic normal form (ANF) of
\begin{equation}
\label{eq:encryption}
g= y \oplus \bigoplus_{i=1}^\alpha \bigoplus_{a=1}^\beta
\bar c_{J(i,a)} \wedge R_{i,a}\;.
\end{equation}
Here $R_{i,a}$ is a
random function. It depends on the same set of 
variables as the clauses $\{c_{J(i,b)} | b\neq a\}$ and is generated in ANF where each possible term occurs with probability $1/2$.

\item The decryption ``algorithm'' is $g(\priv)=y$.
\end{enumerate}
In choosing the clauses in Step~\ref{encryptChoose}, some care is
to be taken:
\begin{enumerate}[label=(\alph*)]
\item \label{item:11} To counter attacks discussed in Section~\ref{sec:const-term-prob}, it
  is preferable if the clauses within one tuple share variables. This
  is particularly important if one of the clauses does not contain
  negations, i.e.\ $s(i,1)=\ldots =s(i,k)=0$.
\item \label{item:12} We have to ensure that each tuple
  shares at least one clause with another tuple to counter the attack
  discussed in Section~\ref{sec:decoding-attack}.
\item \label{item:13} Each
  clause of $\pub$ is to appear in some tuple as discussed in
  Section~\ref{sec:reduced-sat-problem}.
\end{enumerate}

We can ensure \ref{item:12} and \ref{item:13} by choosing $J$ as follows:
First choose a permutation $\sigma \in S_m$ at random. Then set $J(i,a)=\sigma(i+a-1)$, where the
indices are understood modulo $m$. In this scheme $\alpha=m$. 
To also ensure \ref{item:11} one should enhance the probability of
neighbouring clauses, $c_{\sigma(j)}$ and  $c_{\sigma(j)+1}$, to have
variables in common when generating $\sigma$.

To turn the negated clauses in \eqref{eq:encryption} into ANF, one can
use various identities, such as $x\vee y = x \oplus y \oplus xy$, but
$\overline{x\vee y} = \bar x \bar y$ is the most efficient one here. It
leads to
%single
%  \begin{equation}
%  \label{eq:ANFclause}
% 1 \oplus (x_1 \oplus s_1) \vee \ldots \vee (x_k \oplus s_k) 
%  =  (x_1 \oplus s_1 \oplus 1) \wedge \ldots \wedge (x_k \oplus
%  s_k \oplus 1) \;.
%  \end{equation}
%double
\begin{subequations}
 \label{eq:ANFclause}
 \begin{align}
&1 \oplus (x_1 \oplus s_1) \vee \ldots \vee (x_k \oplus s_k) \\
 & =  (x_1 \oplus s_1 \oplus 1) \wedge \ldots \wedge (x_k \oplus
 s_k \oplus 1) \;.
 \end{align}
\end{subequations}
The resulting ANF (obtained by distributing $\wedge$ over $\oplus$) is
of length $ \leq 2^k$. 
The run time to compute the ciphertext in \eqref{eq:encryption}
as well as the length of the resulting ciphertext are in $O(\alpha
2^{\beta k})$. 
Consequently, we need
to choose $\beta \in O(\log(m))$ in order for the run time to be
polynomial.
 The length can be expected to be shorter than the run
time by a factor $>2$ due to the cancellation of terms in the sum and
due to $R_{i,a}$ only
containing $2^{(\beta-1)k-1}$ terms on average.   
If we choose $\beta = \log_2(m)/k$ and $\alpha \propto m$, we
have $\alpha 2^{\beta k} \propto m^2$. For $\beta \in \Theta(1)$, run
time and ciphertext length formally only scale with $m$, but for $m\approx 2^{10}$
and $\beta=k=3$ the pre-factor is still comparable to $m$.
In short, encryption with this setup has complexity $O(m^2)$.

\section{Known Attacks and Improved Schemes\label{sec:known-attacks-impr}}

\yinipar{O}\textsc{bviously}, the most pressing question is whether
the cipher introduced in Section~\ref{sec:single-key} is really
(post-quantum) secure. We will not answer this question, but discuss
some attacks and counters in this Section.
Denote by $G_b =G_b(\pub)$ the set of all possible ciphertexts encoding
$b\in\mathbb B$ and by $G=G_1\cup G_0$ the set of all ciphertexts.
Since the problem of deriving the private from the public key is known
to be hard, we will focus on attacks on the ciphertext. We will refer to
the problem of deciding for $g\in G$ whether $g\in G_0$ or $g\in
G_1$ as the decoding problem. We do not have a proof that this is a
hard problem, although some
related problems are (see Appendix~\ref{sec:some-hard-problems}). In
this section we focus on some particular attacks, i.e.\ algorithms
that solve the decoding problem and show that all of them have
exponential run time, if the parameters of the encryption algorithm
are chosen suitably.

\subsection{Simple Attacks\label{sec:simple-attacks}}

\subsubsection{Enumeration of Ciphertexts
  Attack\label{sec:enumeration-ciphers}}

Since the set of all possible ciphertexts for a given public key is finite, a trivial
brute force attack is to just enumerate it. 
In order to prevent this attack, the encryption parameters $\alpha$ and
$\beta$ need to be large enough. More precisely, the number of
possible ciphertexts is roughly
\begin{align}
\binom{m}{\beta}^\alpha2^{\alpha \beta ((\beta-1)k)}\;.
\end{align}
% Actually, the factor $(\beta-1)k$ should be replaced by the typical
% number of terms appearing in the random functions $R_I$. Anyway, 
We
can ensure this number to be (super) exponentially large by
choosing
\begin{equation}
\label{eq:alBe-strongBound}
 n \in O(\alpha \beta)\;.
\end{equation}
Further we should apply some fixed invertible linear transformation to
the message bit vector before encoding. This way, decoding a few of
the ciphertext bits does not reveal any clear text bits.  
% E.g.\
% $Y_i=\bigoplus_{j<i}y_j$ is in upper triangular form and hence easy to
% invert.

\subsubsection{Constant Term Probability
  Attack\label{sec:const-term-prob}}

Another rather trivial attack is to determine $g(0,\ldots, 0)$ for a
ciphertext $g\in G$. This is just the presence or absence of the constant
$1$ in the ANF, which is where the message bit enters. Hence this
property must not distinguish $G_0$ from $G_1$.

The ANF of a random clause of length $k$ contains a constant $1$ with
probability $1-2^{-k}$, hence the probability for each summand in
\eqref{eq:encryption} to contain the constant $1$ is
$2^{-k-1}$. %, using that $R(0,\ldots,0)=1$ with probability $1/2$.
This is rather small, but 
a variant of the central limit theorem is in this case on our
side,
see Appendix~\ref{sec:boolean-clt}.
%
% the probability that \emph{none} the
% summands in $g$ contain the constant $1$ is
% \begin{equation}
% \label{eq:1}
% \left(1 - 2^{-k-1}\right)^{\alpha\beta}
% \end{equation}
% which becomes exponentially small at the scale of
As a consequence, the probability for $g(0,\ldots, 0)=1$ tends to $1/2$
exponentially quickly with $\alpha\beta$ at the scale of
\begin{equation}
\label{eq:alBe-weakBound}
\alpha\beta \gg  2^{k}
\;.
\end{equation}
Considering the enumeration attack from
Section~\ref{sec:enumeration-ciphers}, the bound 
established in \eqref{eq:alBe-strongBound} is much stronger
than \eqref{eq:alBe-weakBound}. In other words, the
cipher resists the constant term
attack if parameters are chosen such that it resists the 
%most trivial
enumeration attack. 

%\medskip

A more refined version of this attack focuses on those clauses which
can contribute a summand $1$, namely those $c_i$ which do not contain
any negation. For those, 
\begin{equation}
\bar c_i = (x_{i_1}\oplus1)\ldots(x_{i_k}\oplus
1) = 1 \oplus x_{i_1}\oplus \ldots \oplus x_{i_1}x_{i_2}\ldots x_{i_k}
\end{equation}
contributes a constant to the ciphertext sum iff the corresponding $R_i$
also contains the constant $1$.  An attacker can try to judge wether
the latter is the case by looking for the order $k$ term,
$x_{i_1}x_{i_2}\ldots x_{i_k}$, in the ciphertext sum.  The same term
could be caused by another clause depending on the same variables, but
such a clause is part of $\pub$ with vanishing probability for large
$m\propto n$. Hence we need to ensure that $c_i$ shares at least one
variable with another clause in the same tuple or that another clause
in the same tuple contains no negated literals. Both of these can lead
to the appearance of the same order $k$ term, see
the example in Section~\ref{sec:example}.

\subsubsection{Ciphertext Value Probability Attack\label{sec:ciph-value-prob}}

An attacker can try to estimate $|\{g^{-1}(1)\}|$ for any $g\in G$ by
evaluating $g$ on a random set of trial inputs. Hence this value must
not distinguish $G_1$ from $G_0$.  This attack is structurally very
similar to the constant term attack
(Section~\ref{sec:const-term-prob}). For each clause $c$ of length
$k$, the number of inputs on which $c$ is $1$ is $|\{c^{-1}(1)\}|=
2^{k}-1$. The probability of \emph{all} summands in the ciphertext $g$ in
Equation~\eqref{eq:encryption} evaluating to $1$ on a random input is
the same as for none of the summands to contain a constant $1$, if
$R=1$ with probability $1/2$.  Even if the $R$s are chosen
such that they evaluates to $1$ only on $\propto 2^{-\beta k}$ many
inputs, choosing
 \begin{equation}
\label{eq:2}
 \alpha \gg k 
 \end{equation}
still ensures that the probability for $g(1)=1$ is exponentially close
to $1/2$ according to Appendix~\ref{sec:boolean-clt}.
Since $\beta \in O(\log(m))$ for run time
 reasons, condition~\eqref{eq:2} is again ensured by \eqref{eq:alBe-strongBound}.

\subsection{Decoding Attack\label{sec:decoding-attack}}

More sophisticated attacks exploit the structure of
the cipher. 
Any function in $g \in G$ is necessarily of the form
\[
g = y \oplus \bigoplus_{i=1}^m \bar c_i R_i
\]
for some functions $R_i$ (see
Appendix~\ref{sec:structure-ciphers}). 
% This might look as if ciphertexts
% can always be decoded by 
A crucial point of our whole scheme is that polynomial long division
does \emph{not} work over finite fields such as $\mathbb B$, because
the degree is not well behaved under addition and multiplication in
the polynomial ring.  This can be seen in the
Example~\ref{sec:example}.  A simpler example is $x^{n+1}=x$ and hence
e.g.\ $x^2y \oplus y^2 x^5 =0$.  Such ``collisions'' are sufficiently
likely to secure our cipher, if the random functions $R_I$ are chosen
in a good way.

Attacks on the structure of the encoding algorithm start with
the following observation.
The set of all clauses which depend only on literals in a small set
$\{x_{i_1},\ldots, x_{i_{M}}\}$ has an expected size of
\begin{equation}
\binom{M}{k}\frac{m}{n^k}\;,
\end{equation}
which is exponentially small for $M\in o(m)$. In other words, a small
set of clauses $C$ is most likely uniquely determined by $D(C) :=
\bigcup_{c_i\in C}D(c_i)$ where
$D(c(x_1,\ldots,x_k))=\{x_1,\ldots,x_k\}$ denotes the variables on
which $c$ depends. This means that most likely all tuples
$J(i,1),\ldots, J(i,\beta)$ used to
generate our ciphertext $g$ in Section~\ref{sec:encAlg} can be identified
from the ANF of $g$. For a given tuple,
this leaves $2^{\beta(\beta-1)k}$ choices for the random functions
associated with it. 
If we choose $\beta = \log_2(m)/k$ (see Section~\ref{sec:encAlg}) then
$2^{\beta(\beta-1)k} \approx m^{\log(m)}$. This is super polynomial in
$m$, but, for reasonable values of $m$, an attacker can still produce all possible summands associated with
this tuple and try to add them to $g$. He can check whether all terms
involving variables from this tuple cancel. However, the same
variables will occur also in other tuples. In fact, we
should ensure that also some of the \emph{clauses} from this tuple appear in
other tuples. This will leave the attacker with a few possibilities for the random
function, which can only be decided once the touples sharing clauses are
decided as well. But since all tuples are connected
(indirectly) by sharing clauses, the run time of this attack is
exponential in $\alpha$ which we chose to be of the order of $m$.

\subsection{Reduced SAT Problem Attack\label{sec:reduced-sat-problem}}

If an attacker can learn that only a certain subset of clauses was used
in the cipher, he can try to solve the SAT problem given by the
conjunction of only those clauses. If the ratio of
used clauses to variables appearing in those clauses is
small enough, the SAT problem is easily solvable and any solution can be used to decode the ciphertext.

To resist this attack, care is to be taken in Step
(\ref{encryptChoose}) of the encryption
(Section~\ref{sec:encAlg}). The set of all clauses used, $C =
\bigcup_{i=1}^\alpha \bigcup_{a=1}^\beta \{c_{J(i,a)}\}$ should not
depend on more than $|C| n/m$ many different variables.  Furthermore,
reducing the number of clauses used, even at constant clauses to
variables ratio, effectively reduces the key length. Overall, we
should ensure that all clauses are used in the cipher.

\subsection{Oracle Attack\label{sec:oracle-attack}}
The communication according to the algorithms from
Section~\ref{sec:encAlg}  is vulnerable to the following
attack: If the recipient is expected to send a reply to an encrypted
message, an attacker can fake a ciphertext by replacing a literal with a
guessed truth value. If he can learn from the reply whether the ciphertext
was correctly decoded, he gets a strong hint, or even evidence, for the
assignment of this literal in the private key.  Repeating the attack
in a suitable scheme will reveal the full private key.
This is a severe attack which limits the scope of application of the
random cipher to such cases where replies are
only send to authenticated communication partners. Post-quantum
authentication could be provided by e.g.\ a hash-based scheme,
see \cite[Hash-based Digital Signature Schemes]{BBD09}), or by our
identification scheme introduced in Section~\ref{sec:ident-sign}.
One could also use only one-time key pairs for
encryption, again with the problem of authenticating the new keys.

Instead, we have developed two improved version of our scheme. The one
which we consider superior resist the oracle attack completely and is
described in Section~\ref{sec:honest}.  In special circumstances,
also the version described in Appendix~\ref{sec:multi-key} might
be useful. The latter makes the oracle attack substantially more
difficult and preserves the probabilistic nature of the cipher.

\subsubsection{Proof of Honest Encryption\label{sec:honest}}

The version of the
encryption/decryption scheme described here  prevents the oracle attack
completely without increasing the complexity of the encryption
algorithm. Key generation is unchanged, but decryption becomes as
complex as encryption and the feature of a stochastic cipher is lost.
More precisely, using the cipher discussed in
Section~\ref{sec:encAlg}, the ciphertext is not a function of the public
key and clear text. In particular, even if Bob encodes the same clear
text twice, he will get different ciphertexts. This makes
repetition-attacks impossible and the algorithm is very
resistant against rainbow table type attacks. It is, of course,
only pseudo random. In other words, the ciphertext \emph{is} a function of
public key, clear text \emph{and the seed of the pseudo random number
  generator} (PRNG), implicitly used to make the random choices. This
can be used to verify that the ciphertext was not tempered with in order
to oracle the private key, at the cost of loosing the
rainbow-resistance. The latter can then be restored in the usual way
by salting.

Concretely, key generation as described in
Section~\ref{sub:single-keygen} stays untouched and also encryption is
done as explained in Section~\ref{sec:encAlg}, but the sender starts
by seeding the PRNG with a specific seed, computed from clear text and
a salt.  The salt is then to be part of the ciphertext.  
After decryption of the ciphertext (applying it to the private key), the
recipient computes the seed from the clear text and the salt. He then
checks that the received ciphertext text matches that one that is computed
from the public key, the clear text, and the seed according to the
fixed encryption algorithm.  This way, oracle attacks can be detected
without revealing any information. 

The cost to pay is that the
implementation of the PRNG and the encryption algorithm on the sender
and receiver side have to match and that the receiver has to
re-do the most time consuming part of the whole scheme, the
encryption.  Further more, the clear text needs to be long enough (of
order $n$) in order to prevent a brute force attack on the seed.
Notice that the seed, as computed from the salt \emph{and clear text},
is to be considered as a key for decoding the particular message. If
an attacker can find the seed he can decode the ciphertext without
%knowledge of
the private key.

\section{Homomorphic Encryption\label{sec:homom-encrypt}}
 \yinipar{T}\textsc{he} cipher described in Section~\ref{sec:encAlg} is fully
 homomorphic. That means that if Alice wants to know the value of \emph{any} function $f: \mathbb B^n
 \to \mathbb B^n$ on $x \in
 \mathbb B^n$, she can encrypt $x$ bit-wise into the ciphertext
 $c=(c_1,\ldots,c_n) : \mathbb B^n \to \mathbb B^n$ as
 above and then send $c$ to Charley, who has more computation power
 available. Charley computes $f \circ c$ and sends this processed
 ciphertext back to Alice. She decrypts
\begin{align}
\label{eq:homomorph}
\left(f \circ c\right)(\priv) = f(c(\priv)) = f(x) \;.
\end{align}

Notice that oracle attacks by Charley can not be
countered by a proof of honest encryption requirement in this setting (see
Section~\ref{sec:oracle-attack}), or else Alice would have to redo the whole
computation. Anyway, due to the cipher being malleable,
trust in Charly to honestly compute $f$ is needed. 
One might also use multi-Key encryption
as described in Appendix~\ref{sec:multi-key} to check that Charley computes
honestly and discard the key-pair otherwise.

A more severe drawback is that any computation on the ciphertext ``bits''
$c_i$, while not introducing any noise, does increase the length of
the ciphertext. In particular, if $|c_i|$ denotes the number of summands
in the ANF of $c_i$, then 
$|c_i \oplus c_j| \leq |c_i| + |c_i|$ and $|c_i \wedge c_j| \leq |c_i| |c_i|$,
% \begin{align}
% |c_i \oplus c_j| &\leq |c_i| + |c_i|\\
% |c_i \wedge c_j| &\leq |c_i| |c_i|
% \end{align}
where the bounds will likely (almost) saturate for short ciphertexts. As
soon as about half of all possible terms are present, i.e.\ $|c_i|
\approx 2^{n-1}$, the length will likely not grow further. But since
originally $|c_i| \propto m \propto n$,
this means that multiplication leads to an exponential growth in the
ciphertext length. Consequently, only few multiplications are feasible in our scheme.

This growth in length can be traded for ``noise'' by discarding all
terms longer (counting the number of literals in a product) than a
fixed length $L$. The probability of such terms being satisfied by
the (unknown) private key is $2^{-L}$, i.e.\ if only $\ll 2^L$ terms
are discarded, the ciphertext probably still decrypts correctly. The final length
of the ciphertext bits is then limited to $2^L$, but the ``noise''
generated by multiplications will at some point make it impossible to
decrypt the ciphertext, hence effectively also limiting the number of
possible multiplications in $f$.
In this sense our scheme is only somewhat homomorphic and bootstraping
is needed to make it fully homomorphic
without the ciphertexts becoming unfeasibly large.

\section{Identification and Signatures\label{sec:ident-sign}}

\yinipar{I}\textsc{n} this sections we introduce a zero-knowledge proof for our key
pairs. This leads to an identification and signature scheme, which is
independent from the encryption scheme discussed above, apart from
sharing the same type of public/private key pairs. In fact, the scheme
proposed here is a variant of the one by Blum based on the Hamitlon
cycle problem \cite{Blum1986}. The latter is NP-complete, hence
polynomial-time equivalent to SAT, but one can apply the ideas more
directly.

% \cite[Blum1991] suggest a zero-knowledge proof for $3$-SAT, which,
% however, seems to be based on a variant of discrete logarithms and
% hence is not post-quantum secure.

\subsection{Zero-Knowledge Proof/Identification\label{sec:zero-knowl-proof}}

Let $\pub$ again be a $k$-SAT instance in $n$ boolean
variables and $\priv \in \pub^{-1}(1)$ a solution only known to
Alice. The zero-knowledge proof consists of
$K$ rounds, each of which proceeds as follows:
\begin{enumerate}
\item Alice chooses a random invertible function $f : \mathbb B^n \to
  \mathbb B^n$ and commits to $f^*(\pub) = \pub \circ f$ and to
  $f^{-1}(\priv)$. More precisely, the commitment to $f^*(\pub)$ is to
  be in the form of committing to each literal in the pull back of the
  form, i.e.\ to each $f^*(x_I(1,1)), f^*(x_I(1,2)), \ldots,
    f^*(x_I(m,k))$.

\item \label{item:14} Bob chooses whether Alice should reveal either
\end{enumerate}
\begin{enumerate}[label=2.\alph*)]
\item\label{item:15} all of $f^*(\pub)$ or
\item\label{item:16} $f^{-1}(\priv)$ together with $\left\{f^*(x_{I(i,a_i)}) | i \in
    \{1,\ldots,m\}\right\}$, where Alice randomly chooses the $a_i \in
  \{1,\ldots,k\}$ such that
  $f^*(x_{I(i,a_i)})\left(f^{-1}(\priv)\right)=1$ for all $i$. 
\end{enumerate}
In this scheme Bob either verifies that Alice did not cheat when
generating $f^*(\pub)$ or that she indeed knows a solution. Fake
proofs will hence be discovered with probability $1/2$ each
round and the probability to get through all rounds is 
$2^{-K}$.

The run time of the above scheme depends on what kind of functions we
allow for $f$. If we only use permutations\footnote{Here the
  permutation acts on $\mathbb B^n$ by permuting the coefficients with
  respect to the fixed base and correspondingly on formulas by
  permuting the indices of the variables.}, the run time for
generating the permutation, its inverse and reshuffling $\priv$ is
$O(n)$ and $O(m)$ for re-labelling $\pub$. However, in this case the
number of $1$s and $0$s in the public key is leaked by revealing
$f^{-1}(\priv)$. 
% This still leaves $\binom{n}{N}$ possibilities for
% $\priv$, where $N$ is the number of $1$s (or $0$s). Generically
% $N\approx n/2$, in which case this corresponds to only about
% $\SI{2}{\bit}$ of information about $\priv$. If, however, $N$ is
% accidentally very large/small this leakage could substantially weaken
% the key pair.  
To avoid this, we should at least add a random affine shift
$s\in\mathbb B^n$ to the permutation, i.e.\ $f(x)=\sigma(x) + s$.
% If we allow for any linear (or even higher order)
% function, the above run times and resulting length of $f^*(\pub)$
% might get much longer. To avoid this, we might use linear functions
% with band-matrix form, i.e.\ $f^*(x_i) = \sum_j f_{i,j}x_j$ with
% $f_{i,j}=0$ for all $i-j>b$, where $b$ is called the band (half)
% width. If $f^*(x_i)$ is given in ANF, it has length $\leq 2(b + 1)$.
% We can also precompose such a random invertible band matrix with a
% random permutation to get a more general class of linear invertible
% functions. Notice that $f^{-1}$ will in either case \emph{not} be a
% band matrix, generically, and we should, of course, only compute
% $f^{-1}(\priv)$ rather than all of $f^{-1}$. For tri-diagonal matrices
% ($b=1$) we can still generate $f$ and $f^{-1}(\priv)$ in $O(n)$ and
% re-label in $O(m)$.

\subsection{Signatures\label{sec:signatures}}

By a Fiat-Shamir heuristic we can convert the above interactive
identification into a signature scheme. To this end, let $h: \mathbb
B^* \to \mathbb B^K$ be a cryptographic hash function. To sign a document:
\begin{enumerate}
\item Alice generates $K$ random functions, $f_1,\ldots,f_K$, as in
  Section~\ref{sec:zero-knowl-proof} and commits to all 
  $f_i^{-1}(\priv)$ and $f^*(\pub)$.
\item She then computes a hash of the document to be signed,
  concatenated with (a suitable encoding of) her commitments. Denote
  the latter by $C$.
\item She reveals the information as in Step~\ref{item:14} of
    Section~\ref{sec:zero-knowl-proof}, choosing \ref{item:15} or
      \ref{item:16} in the $i$th round if the $i$th bit of the hash is
      $0$ or $1$, respectively. Denote this
        revealed information by $R$.

\end{enumerate}
The signature consists of $C$ and $R$. It is verified by Bob by checking that the correct
information was revealed by re-computing the hash. He also checks that
the revealed information is valid as in
Section~\ref{sec:zero-knowl-proof}, of course.

To fake this signature, Alice could fake the zero-knowledge proofs and
generate signatures until, by chance, a valid one is
generated. However, the probability for this to happen is as small as
cheating in the identification scheme, i.e.\ exponentially small in
the number of rounds.

\section{Conclusion\label{sec:conclusion}}
\yinipar{T}\textsc{he} motivation for developing the public key
crypto-scheme based on the Boolean Satisfiability Problem discussed in
this paper is to fill the arsenal of post-quantum cryptography with
some fresh ammunition. The simplicity of the scheme developed in this
paper might be an advantage over the well known post-quantum
schemes. More conceptually, we are not aware of any scheme which uses
random ciphertexts in a similar way (compare to
Figure~\ref{fig:encoding}).  Furthermore, there is quite some freedom
in the details of the cipher, in particular in choosing
probability distributions for the random functions $R_I$. This makes
it possible to adapt the algorithm, if more sophisticated attacks are
discovered in the future.  The cipher presented in
Section~\ref{sec:encAlg} has undergone some evolution to resist all
attacks which came to our mind. The version using a proof honest
encryption has no vulnerabilities currently known to us, but of course
much more crypto-analysis is needed and the reader is invited to
devise stronger attacks to challenge and improve the algorithm. In
particular, we have not proven that it is (NP-)hard to decipher a
message without knowledge of the private key.  It is the problem of
deducing the private from the public key that is NP-hard, i.e.\
``post-quantum'', by construction.  The same is true for the signature
scheme presented in Section~\ref{sec:signatures}, which is independent
of our encryption scheme. It is analogous to Blum's well known scheme
\cite{Blum1986}, adapted to our SAT scenario.

A notable feature of our cipher is that, in principle, it is fully
homomorpic, i.e.\ applying any function to the ciphertext bit vector and
then decoding yields the same result as decoding and then applying the
function. However, as discussed in Section~\ref{sec:homom-encrypt},
the ciphertext might become unfeasibly long if to many multiplications are
applied to it, hence our scheme is effectively only somewhat
homomorphic.  The oracle attack mentioned in
Section~\ref{sec:oracle-attack} is a stronger incarnation of
malleability. It is a severe generic attack on
any cipher consisting of Boolean functions.  Enforcing honest
encryption, as explained in Section~\ref{sec:honest}, is a generic
counter.  The multi-key version of our scheme described in
Appendix~\ref{sec:multi-key} is another work
around, however, not completely resistant. Although it only reveals much less information, the key pairs
still need to be changed regularly, but here this could here be feasible. In
some special situations, like using our scheme for homomorphic
encryption, the multi-key version might be
preferable. Encryption is considerably more complex than key
generation and decryption, which in particular yields some protection
against DOS attacks for multi-key schemes.

The length of the public key and the run time of the key
generation algorithm scale as $O(n \log n)$ with the length of the
private key $n$. The length of the ciphertext and the run time of the
encryption algorithm per bit of clear text scale as
$O(m^{1+\epsilon})$ with $\epsilon\geq 0$ and we have identified
$2\alpha_c(k) n > m
> \alpha_c(k) n$ as the relevant parameter range for hard planted SAT
instances (see Appendix~\ref{sec:benchmarks}). The crucial
question is how big $n$ should be in practice today.  At the SAT
Competition 2014\footnote{\url{http://www.satcompetition.org/2014/}},
random $3$-SAT instances of size $n\approx 10^4$ have been solved for
$m\approx m_c$ in less than one
hour.\footnote{E.g.~\url{http://satcompetition.org/edacc/sc14/experiment/24/result/?id=23531}}
These instances were selected in order to be solvable within that
time, but still they are randomly generated with non-negligible
probability.  
% For $m<m_c$ even instances with $n\approx 10^6$ have
% been solved, but these instances have most likely very many solutions
% and are hence much easier than our public keys.  
Our own benchmarks
(Appendix~\ref{sec:benchmarks}) show that the MiniSat
solver \cite{minisat} can not break keys of
length $n > 2^{10}\approx 10^3$ on a modern PC using one
thread. Taking large scale parallelisation into account, we should at
least choose $n > 2^{11}$, but in view of the SAT
Competition results, $n > 2^{14}$ seems more advisable.

Key generation is very fast and choosing even $n > 10^7$ is not a
problem here. However, encryption with our not very much optimised
proof of concept implementation \cite{code} already takes time of the
order of few seconds per bit of clear text to encode with $\alpha=m=5n
= 5 \times 2^{10}$ and $\beta=3$ and still tenth of seconds per bit
for $\beta=2$ and the other parameters as before.  Although there is
certainly room for improving the implementation, the constraint
$\beta\geq 2$ means that the run time of the encryption algorithm is
in $O(m^{1+\epsilon})$ with $\epsilon \geq 3/5$ for $m\approx 2^{10}$.
These run times are not yet very well suited for practical
applications.  However, the situation could be improved by advancing
away from bit-wise encryption.  Encoding longer messages at once is an
interesting topic for future research. One could use the (vector
space) embedding of polynomials with binary coefficients into
polynomials with coefficients in any finite field $\mathbb F_q$ to
encode the message in a shorter $q$-adic representation decimal by
decimal. However, increasing $q$ makes polynomial long division more
likely possible, hence the security considerations in
Section~\ref{sec:decoding-attack} need to be seriously reviewed in
this setting.  If one can counter the attack from
Section~\ref{sec:reduced-sat-problem} by other means than using all
clauses, smaller $\alpha$ would also lead to a substantial speed
up. Anyway, before more research builds on the ideas presented here,
we would like our algorithm to be challenged by more advanced attacks
in order to improve it. Furthermore, the problems which we have proven
to be NP-hard (see Appendix~\ref{sec:some-hard-problems}) are still
not very close to the decoding problem. To build faith in this
cipher really being post-quantum secure, more research in this
direction is in order.

\section*{Acknowledgement}
We acknowledge helpful discussions with Rainer Schrader and
Ewald Speckenmeyer as well as their research groups in Cologne. In
particular, we would like to thank Martin Olschewski, who pointed out
the oracle attack.  

We are very thankful to Paulo Barreto for friendly advice and his
interest in our first draft of this paper. In particular his
suggestion to introduce signatures via a zero knowledge proof brought
this idea home to us and he stressed that being homomorphic might be an
important property of our scheme, rather than a side effect.

% We thank Ricardo Kennedy for urging us to produce a picture (leading
% to Figure~\ref{fig:encoding} and \ref{fig:decoding}).

The author was supported by Martin Zirnbauer, by a grant
from Deutsche Telekom Stiftung, and another grant from the
Bonn-Cologne Graduate School of Physics and Astronomy, funded by the
DFG.

%  \bibliographystyle{plain}
%  \bibliography{Bib}{}

\begin{thebibliography}{1}

\bibitem{BBD09}
D.~J. Bernstein, J.~Buchmann, and E.~Dahmen, editors.
\newblock {\em Post-Quantum Cryptography}.
\newblock Springer-Verlag Berlin Heidelberg, 2009.

\bibitem{Blum1986}
M.~Blum.
\newblock How to prove a theorem so no one else can claim it.
\newblock {\em ICM Proceedings}, 2:1444--1451, 1986.

\bibitem{minisat}
N.~Een and N.~Sorensson.
\newblock Minisat.
\newblock \url{http://www.minisat.se/}.

\bibitem{Kirkpatrick1994}
S.~Kirkpatrick and B.~Selman.
\newblock Critical behavior in the satisfiability of random boolean
  expressions.
\newblock {\em Science (New York, N.Y.)}, 264(5163):1297--1301, 1994.

\bibitem{Krzakala2012}
F.~Krzakala, M.~Mezard, and L.~Zdeborova.
\newblock Reweighted belief propagation and quiet planting for random k-{SAT}.
\newblock {\em Journal on Satisfiability, Boolean Modeling and Computation},
  8:149--171, 2014.

\bibitem{code}
S.~E. Schmittner.
\newblock A prove of concept implementation of a {SAT}-based encryption scheme.
\newblock \url{https://github.com/Echsecutor/kryptoSAT}, 2015.

\end{thebibliography}

\clearpage
\appendices%weird ieee appendix format

\section{Some Hard Problems\label{sec:some-hard-problems}}

In this section, we establish that some problems related to the
decoding problem (see Section~\ref{sec:known-attacks-impr}) are hard.
To this end, fix a public key $\pub$ and assume that
$\pub^{-1}(1)=\{\priv\}$. The partitioning of ciphertexts $G=G_0\cup G_1$ is characterised by
\begin{align}
G_1 &= \{ g \in G \; | \; \pub \Rightarrow g \}\\
G_0 &= \{ g \in G \; | \; \pub \Rightarrow \bar g \} \;.
\end{align}

Deriving the private key is harder than decoding, since $\forall g\in
G : g\in G_{g(\priv)}$. Assuming that all $g \in G$ are given in
such a form that evaluating $g(\priv)$ is possible in polynomial time, the decoding problem is in NP.

If $G$ would contain the elementary functions $X:=\{x \mapsto x_i\}$
then the decoding problem would be (polynomial time) equivalent to
deriving the private key, hence solving the SAT problem.  A set of
functions $Y$ of polynomials size will be called ``hard to decode'',
if solving the decoding problem for each $f\in Y$ determines (by a
polynomial time algorithm) the solution of the SAT problem. 
The following
sets of functions are hard to decode:
\begin{enumerate}
\item\label{item:6} The elementary functions $X$

\item\label{item:7} Any set of functions containing a subset that is hard to decode

\item\label{item:2}
  $\{f \oplus s_f | f \in Y\}$ where $s_f \in \mathbb B$ are known and
  $Y$ is
  hard to decode

\item\label{item:1}
 $\{f \oplus g | f \in Y\}$ where
 $Y$ is hard to decode and $g$ is any Boolean function 

\item\label{item:3}  $\{f \wedge g | f \in Y\} \cup \{f \wedge \bar g | f \in Y\}$
  with $Y$, $g$ as above

\item\label{item:4} $\{x_i \wedge f | f \in Y, x_i \in X\}$ with $X$, $Y$ as above 

\item\label{item:5} $\{x_i \vee f | f \in Y, x_i \in X\}$ with $X$, $Y$ as above 

\end{enumerate}
(\ref{item:2}) is hard to decode since $f \oplus s \in G_b
\leftrightarrow f \in G_{b\oplus s}$.
(\ref{item:1}) is hard to decode since we can decide for each $f \oplus
g$ whether or not it is implied by $\pub$, then assume that
$g(\priv)=0$ and use that $Y$ is hard to decode to produce a trial
solution $\priv'$. If $\pub(\priv')\neq 1$ then $g(\priv)=1$ and we
can use (\ref{item:2}). (\ref{item:3}) is hard to decide since one of
the two sets evaluates to $\{0\}$ on $\priv$. Since there are only two
possibilities for $g(\priv)$ one can check both in polynomial time,
similar to (\ref{item:1}). In (\ref{item:4}) (and similarly in (\ref{item:5})) we can
decide $\{x_1 \wedge f\}$. If this turns out to be a subset of $G_0$
then we likely have $x_1(\priv)=0$ and we proceed deciding $\{x_2
\wedge f\}$. After polynomial time we either arrive at some $x_i(\priv)=1$
and can hence derive $\priv$ or conclude that $\priv =(0,\ldots,0)$.

The above construction shows that some generic problems similar to the
decoding problem are hard to decide. Notice, however, that the sender
can not (on purpose) encode a message of polynomial length into a hard
to encode set of functions in polynomial time, or else he would solve
the SAT problem.

\section{Boolean Central Limit Theorem\label{sec:boolean-clt}}

Consider a set of i.i.d.\ Boolean variables $x_i$, with
$p_1:=\operatorname{prob}(x_i=1)$. Then
\begin{align}
p^\oplus_0:=\operatorname{prob}\left(\bigoplus_{i=1}^M x_i=0\right) &=
\sum_{j=0}^{\lfloor M/2 \rfloor} \binom{M}{2j}p_1^{2j}(1-p_1)^{M-2j}
\\
p^\oplus_1
:=\operatorname{prob}\left(\bigoplus_{i=1}^M x_i=1\right) &=
\sum_{j=0}^{\lceil M/2 \rceil-1} \binom{M}{2j+1}p_1^{2j+1}(1-p_1)^{M-2j-1}
\end{align}
and hence $|p^\oplus_0 - p^\oplus_1| = |1-2p_1|^M$ converges to $0$
for any $0<p_1<1$ and $M\to\infty$ at exponential speed. In other
words, for large enough $M$ we have $p^\oplus_1 \approx 1/2$
irrespective of $p_1$. Large enough here means $M \gg -1/\log
|1-2p_1|$, which means $M \gg p_1^{-1}/2$ for small $p_1 \ll 1/2$.

%M ln(1-2p_1) << -1 <=> M >> -1/ln(1-2p_1)

\section{Structure of Ciphertexts\label{sec:structure-ciphers}}

In this section we discuss which functions $g$ can, in principle, be constructed from
the public key $\pub$  without knowledge of the
private key, such that $\pub \Rightarrow g$, i.e. $g\in G_1$. We
will therefore assume that only the clauses $c_i$ of $\pub =
\bigwedge_{i=1}^m c_i$ can be used as the elementary building blocks
for which $\pub \Rightarrow c_i$ is known.

Any encryption algorithm of the type discussed in this paper will lead
to sets of ciphertexts $G$ and $G_b$ which are contained in the sets
$\tilde G$ and $\tilde G_b$, respectively. The latter are constructed
as follows
\begin{enumerate}
\item\label{item:8} $1 \in \tilde G_1$ and $c_i \in \tilde G_1$ for $i\in\{1,\ldots,m\}$.
\item\label{item:9} $f\in \tilde G_a, g\in \tilde G_b \Rightarrow f \oplus g \in
  \tilde G_{a\oplus b}$, $f \wedge g \in \tilde G_{a\wedge b}$ and $f \vee g \in
  \tilde G_{a\vee b}$. In
  particular, $\bar g = 1 \oplus g \in \tilde G_{\bar b}$.
\item\label{item:10} $f\in \tilde G_1, g$ arbitrary $\Rightarrow f\vee g
  \in \tilde G_1$ and $\bar f \wedge g \in \tilde G_0$.
\end{enumerate}

The sets that can be constructed from
(\ref{item:8}) using (\ref{item:9}) and (\ref{item:10}) are
\begin{align}
\tilde G_0 &= \left\{\bigwedge_{i=1}^m \bar
c_i f_i \; | \; f_i : \mathbb B^n \to \mathbb B\right\}\\
\tilde G_1 &= \{1 \oplus f\; | \; f \in G_0\}
\;.
\end{align}
These are stable under (\ref{item:9}),
since $f \vee g = 1 \oplus \bar f \bar g = f \oplus g \oplus fg$ and
$f = 1 \oplus \bar f$. Further, the set of Boolean functions used in
(\ref{item:9}) is complete.

\section{Multi-key Chains\label{sec:multi-key}}

In some situations, the proof of honest encryption method laid out in
Section~\ref{sec:oracle-attack} might not be favourable, for example if the
feature of a (pseudo) random cipher is crucial or re-computing the
encryption on the receiver side is to costly. In such cases, we can
still substantially weaken the oracle attack.  The key idea here is to
introduce redundancy, which increases key length and all run times by
a constant factor.

The private/public key chain now consists of
$\gamma \in \mathbb N$ different private/public key pairs. To encrypt a message,
each bit is to be encrypted with each of the public keys, i.e.\
the ciphertext for one bit now consists of $\gamma$ many Boolean functions
in ANF. In a valid ciphertext, all of these functions evaluate to the same
value upon inserting the respective private keys.

After decoding, the recipient has to decide whether or not to accept
the message (bit), if the ciphertext is invalid. Rejecting all invalid
ciphertexts would reveal as much information about the private key as the
simple version of Section~\ref{sec:single-key}. Therefore we fix a
threshold $t \in \{2,\ldots,\gamma/2\}$ and accept the bit with value
given by the majority, if the minority is smaller than $t$. To
properly choose $t$, we consider the two possible outcomes of the
attack:
\begin{enumerate}
\item If the message is rejected, the attacker learns that more than
  the (known) threshold $t$ of the bits he guessed did not match the
  private keys. \label{it:rej}

\item If the message is accepted, the attacker will learn whether or
  not he guessed the majority of bits right (from the reply of the
  receiver).

\end{enumerate}

To keep the information leakage about the private key as small as
possible, we should choose $t$ large in order to make \eqref{it:rej}
unlikely. More precisely, if $f\in \{0.\ldots,\gamma\}$ of the
encoded versions of the bit are manipulated by replacing one variable with a
guessed value (such that the change in the function influences
its value), the probability of rejection is
\begin{align}
\operatorname{prob}_r(t,f)= \frac{1}{2^f} \sum_{t\leq T\leq f}\binom{f}{T} \;,
\end{align}
which can be expressed through the error function for large $f$.
If $f$ is close or even equal to $t$, the attacker gains substantial
information, but for this attack, the success probability is
exponentially small in $t$. 

In the opposite case of choosing
$f$ close to $\gamma$, the attack is most likely detected if $t$ is
small enough. Concretely choosing e.g.
\begin{equation}
t = \frac{\gamma}{2} - \frac{c}{2}\sqrt{\gamma}
\end{equation}
with $c=3$ means that more than $99.7\%$ (for large $\gamma$) of the attacks will be
detected with hardly any information gain for the attacker. More
precisely, the probability of a successfully attack of this type
scales down super exponentially (with the complementary error
function) and is below $10^{-8}$ already for $c=6$. In practice,
choosing $t\propto\gamma$ with a proportionality factor slightly
smaller than $1/2$ should be most reasonable. 

The attacker gains most information from an attack with
$\operatorname{prob}_r(t,f)\approx 1/2$, i.e.\ $f \approx 2 t$. This
attack being rejected or not indicates that more or less than half of
the private key bits were guessed correctly. Gathering this knowledge
from $O(\gamma^2)$ attacks in a suitable scheme will reveal the value
of all $\gamma$ bits of the keys with a high confidence level on the
attackers side. Hence the multi-key hardened version is not fully
resistant against the oracle attack, but the attack will almost surely
be noticed before substantial information is leaked. This invalidates
the authenticity of the attacker in an authenticated
communication. Even if information was leaked, not all keys in the
chain have to be replaced, which limits the damage of a successful
attack.  

One can improve the scheme a little more by also choosing
$\gamma$ at random for each decryption, but still statistical analysis
will eventually reveal the private key bits. Hence the proof of honest
encryption scheme
introduced in Section~\ref{sec:honest} is certainly more secure than
the multi-key scheme.

\section{Benchmarks\label{sec:benchmarks}}

We have conducted some simple bench marks for breaking the public key,
i.e.\ solving the SAT instance, using
MiniSat \cite{minisat}.  It is well known that
random instances are the hardest for $m\approx m_c = \alpha_c(k) n$. For the planted
instances that we use as public keys, it turns out that the hardest
instances have slightly more clauses. 
% Still, using too large $m/n$
% makes it easier to find the solution.

\subsection{3-SAT}

For $k=3$ we find (see Figure~\ref{fig:k3r8} to \ref{fig:k3r4.3})
that the hardest instance have $m \approx 5 n$ where $m_c \approx 4.2
n$ \cite{Kirkpatrick1994}.  In Figure~\ref{fig:k3r4.3} we show the
run time of MiniSat for various random instances generated by our key
generator as a function of the number of variables for fixed
$m/n=4.3$. This is very close to the critical ratio. Fitting an
exponential function to the minimal run times, we extrapolate that one
should choose $n_{100}\approx 1500$ to ensure that the MiniSat solver
would take at least $100$ years to break the public key. For $m/n=4.5$,
as displayed in Figure~\ref{fig:k3r4.5}, we find the same $n_{100}$
within the error of our approximation, but for $m/n=5$
(Figure~\ref{fig:k3r5}) $n_{100} < 1000$ is significantly
smaller.  Increasing $m/n$ further leads to an increases in $n_{100}$ which
again reaches $n_{100}\approx 1500$ for $m/n=8$.
Consequently, we choose $m/n=5$ and a private key length
$n=2^{10}$ for $k=3$ as the default values for the 
 proof-of-concept implementation \cite{code}. 
This leads to a public key length of about $k * m
*(\log_2(n)+1) \approx \SI{165}{\kilo\bit}$. Notice that the keys
generated with these parameters can be considered ``PC-secure'' but
securing the keys against more sophisticated massively parallel
solvers rather requires $n \gtrsim 2^{11}$.

% Encoding $\SI{16}{\bit}$ with such a key with 

\subsection{4-SAT}

For $k=4$ the critical ratio of clauses to variables is about
$m_c=9.8n$. Using $m/n=10$ our MiniSat benchmarks indicate
$n_{100}\approx 350$. So the private key size can be reduced
significantly by using higher $k$. However, the public key size for these
parameters ($\approx \SI{133}{\kilo\bit}$) is comparable to the one for $k=3$.
This means that the run time for encryption with higher $k$ is significantly longer
(exponential in $k$, see Section~\ref{sec:encAlg}).
Overall, it does not pay off to use higher values of $k$.

\input{plots}

\end{document}